

\magnification =1200
\baselineskip 0.6cm

\def\i{\imath }
\def\j{\jmath }

\def\12{{\scriptstyle {1\over 2}}}

\def\G/H{\scriptscriptstyle G/H}

\def\L'{\scriptscriptstyle \Lambda '}
\def\l'{\scriptscriptstyle  \lambda '}

\def\pl{P^{\scriptscriptstyle \lambda ,\lambda '}}

\def\lb{{\bar \lambda}}

\def\lbm{({\lambda .\beta \over {\mu }})}

\hfill {BONN-HE-93-1}

\hfill {SUTDP/93/71/1}

\vskip 1.4cm

\centerline {\bf MODULAR INVNT PARTITION FUNCTIONS}

\centerline {\bf AND METHOD OF SHIFT VECTOR }

\vskip 1.2cm
\centerline {{\it{A.Shirzad }}\footnote*
{Research supported by
Sharif University of Technology}}

\centerline { Department of Physics, Sharif University of Technology }
\centerline { P.O.Box 11365-9161, Tehran, Iran }
\vskip 1cm

 \centerline {{\it H.Arfaei }\footnote{**}
{On leave from Sharif University of Technology, Tehran Iran}}

\centerline {Physics Institute ,Bonn University}
\centerline {Nussalle 12 , W-5300 Bonn 1, Germany}
\centerline {and}
\centerline {Max-Planck Institute for Mathematics,}
\centerline {26 Gottfried-Claren  Strasse   W-5300 Bonn 3,Germany}

\vskip 1.5cm

\centerline{ Abstract }
Using shift vector method we obtain a large class of self-dual lattices  of
dimension $(l,l)$, which has a one to one correspondence with modular
invariants of free bosonic theory compactified on co-root lattice of
a rank $l$ Lie group. Then a large number of modular invariants of
affine Lie algebras are derived explicitly. As two applications of this method,
we give a direct derivation of $D$-series of $SU(N)$ and a new proof
for the A-D-E classification of the $SU(2)_k$ partition functions.

\vfill \break

{\bf{1-Introduction}}:

It is well known for long time that modular invariance of the partition
functions strongly restricts the possible conformal field theories constructed
based on
the affine lie algebras of currents [1,2]. Much work is done to find all the
modular invariants and all modular invariant partition functions of these
theories [3,4,5,6,7]
Despite the amount of work only partial success is achieved. The problem is
solved for the $SU(2)_k $ case  completely [7,8]  resulting in the so called
A-D-E classification. The case of other affine algebras is not settled;  in the
$SU(3)$ case few exceptional partition functions are obtained [9,10] apart from
 the D series only few cases are found for $SU(N)$. In these cases there is no
proof of the completeness [6,10].

Several methods are devised to construct modular invariants among which we
can name conformal embeddings [3] , orbifold constructions [4], simple currents
[5] , shift vectors [11] ,and direct search for the commutants of the
generators of the modular transformations. In this article we apply the shift
vector method and use a mapping of the modular invariants of the bosonic
theories to that of WZNW models to make progress in this program.
Large class of partition functions are obtained in this way including the
$SU(2)$
solutions and  the $SU(N)$ D-series. Based on this method we also give a new
proof of the A-D-E classification. The $SU(3)$ exceptional solutions are also
found but due to their lengthy discussion we defer it to another publication.

. When current algebra is the
symmetry
of a theory in both the left and right sectors, [2] the Hilbert space
consists of combinations of its left and right integrable representations ,
 $${\cal H}=\oplus_{\Lambda ,\Lambda '} M^{\Lambda ,\Lambda '}
H_{\Lambda}\otimes {\bar H}_{\Lambda '} \eqno (1-1)$$
where $M^{\Lambda ,\Lambda '}$ is the multiplicity of the representation
labeled by highest weights ${\Lambda}$ and ${\Lambda '}$ in the left and right
sectors respectively.
The partition function of the theory is: $${\cal Z}(\tau )=\sum _{\Lambda
,\Lambda '}
M^{\Lambda ,\Lambda '}\chi_{\Lambda }(\tau )
{\bar \chi}_{\Lambda '}({\bar \tau })\eqno (1-2)$$
where $ \chi_{\Lambda }(\tau )$ and $ {\bar \chi}_{\Lambda '}({\bar \tau }) $
are characters of representations $ H_{\Lambda} $ and $ \bar H_{\Lambda '}$
respectively.

For  ${\cal Z}(\tau )$ to be invariant
under the action of modular group on the torus
parameter,  $M^{\Lambda ,\Lambda '}$ as a matrix  acting  on the finite set
of grable representations is subject to two important restrictions : first it
should have
non-negative integers as its elements; and second, it should commute with the
representation of the modular group generators $S$ and $T$, on characters of
the
given Kac-Moody algebra. Moreover the multiplicity of vacuum should be
one and primary fields of the theory should be closed under the operator
product algebra. Finding the set  of consistent WZNW theories corresponds to
finding all possible M's for a given affine $G$ (represented by $\hat G$) at
arbitrary level
$k$.
To find this M's we use a correspondence with the modular invariants of a
subgroup H of G having the same rank [11].

 Suppose that
 $$ \sum P^{\lambda ,\lambda '}\chi ^{(H)}_{\lambda }
\bar {\chi}^{(H)}_{\lambda '}\eqno (1-3)$$
is a modular invariant of $\hat H$ at level $(k+{\tilde h}_G-{\tilde
h}_H)$.Then
$$\sum M^{\Lambda ,\Lambda '}\chi ^{(G)}_{\Lambda }
\bar \chi^{(G)}_{\Lambda '}\eqno (1-4)$$
is   a modular invariant of $\hat G$ at level $k$ provided that
$$\sum M^{\Lambda ,\Lambda '}=P^{\lambda ,\lambda '}
 I_{\lambda }^{\Lambda } I_{\lambda '}^{\Lambda '}\eqno (1-5)$$
Here,${\tilde h}_G$ and ${\tilde h}_H$ are dual Coxeter numbers of $G$ and $H$
respectively,.  $I_{\lambda }^{\Lambda }$ and $I_{\lambda '}^{\Lambda '}$
are the  {\it{supersymmetry indices}} [11] which can be obtained in the
following way:

If there exist elements $ \omega \in   {W(G)\over W(H)}$ and
$ \mu \in {\Gamma (G)\over \Gamma (H)}$
such that $$(\lambda +\rho_H)=\omega (\Lambda +\rho_G)+(k+\tilde h_G)\mu ,\eqno
(1-6)$$
then we have: $$I_{\lambda}^{\Lambda}=\epsilon (\omega );\eqno (1-7)$$
otherwise
$I_{\lambda }^{\Lambda }$ vanishes. Here $W(G)$is the Weyl group, $\Gamma (G)$
is the co-root lattice and $\rho_G$ is half of the positive roots of $G$;
 similar definitions hold for $H$, and $\epsilon (\omega )$ is determinant
of the Weyl transformation $\omega $.
We can see that  for  each  $\Lambda $ the number of
corresponding $\lambda $'s with nonvanishing $I_{\lambda }^{\Lambda }$ is
: $$p=\Bigg \vert {{W(G)\over W(H)}}\Bigg \vert \Bigg \vert{\Gamma
(G)\over
\Gamma (H)}\Bigg \vert .\eqno (1-8)$$

Moreover $I_{\lambda }^{\Lambda }$'s satisfy the following identity :
$$\sum _{\lambda }I_{\lambda }^{\Lambda } I_{\lambda }^{\Lambda '}
=p\delta ^{\Lambda \Lambda '}\eqno (1-9)$$

Warner [11] has  shown that every modular invariant of $\hat G$ can be obtained
from a modular invariant of $\hat H$, although several different modular
invariants of
$\hat H$ can lead to the same invariant for $\hat G$.

One special choice for $H$ is the maximal torus of $G$, that is:
$$H=[U(1)]^l\eqno (1-10)$$
where $l$ is the rank of $G$. As we will  discuss in section 2, the problem
of finding all modular invariants of $H=[U(1)^l$  (and consequently of $G$ )
is
reduced to finding all even Lorentzian self-dual lattices [13] with signature
$(l,l)$. For this reason, we have exploited the  shift vector method in order
to
construct new self-dual lattices from some given self-dual lattice.

The structure of this paper is as follows:
In section (2) after reviewing some aspects of compactified bosons, we will
find an explicit form for $I_{\lambda }^{\Lambda }$, consequently the relation
(1-5)
between modular invariants of $\hat H$ and $\hat G$ would have a simple and
clear form. Then an explicit expression
for $P^{\lambda ,\lambda '}$, which is the modular invariant resulted from
 employing a given shift on diagonal
lattice, would be found. Particularly, we will find which characters in the
right hand sector would be coupled to a given character in the left hand
sector.

It should be mentioned that in order to find all Lorentzian even self-dual
lattices (and hence all modular invaria$P^{\lambda ,\lambda '}$ ),
beginning with the diagonal
lattice, we should iterate the procedure of shifting on each new lattice
we find. But in this paper we  consider only first shifting of diagonal
lattice.
 We call this process $mono-shift$, in contrast with $multi-shift$
which refers to the complete iterating process. Fortunately mono-shift method
in
its own
right is rich enough, and a vast number of known  modular invariants are
consequences of mono-shifts only.

In sections 3 and 4 we present two important applications of our method.
The first one is derivation of the $D$-series of $SU(N)$ at arbitrary level
$k$,
by
choosing a particular type of shift vectors . As the second application,
we give a complete discussion about the modular invariants of $SU(2)$ . We will
show that in
this special case, mono-shift method
 provides a new  proof for the so-called
$A-D-E$ classification.  In the last section, we shall discuss         the
advantage
and difficulties of this method
when we attempt to  generalize it to other Lie groups like $SU(3)$. Our
results
on higher rank algebras is   deferred                   to another publication
due to its length

\vskip 10pt

{\bf{2-The Monoshift Method}}

\vskip 5pt

In the following  we choose $H$ to be an abealian subgroup of $G$,
we can set $\Gamma (H)=\Gamma (G)$, and consider the $\hat H$-theory as $l$
free
bosons compactified on the integer lattice $\sqrt {g}\Gamma (G)$ [7], where
$$g=k+\tilde {h}_G\eqno (2-1)$$
is known as the $height$
 for $\hat G$ at level$k$, which is also considered as the level for
$\hat H$-theory (since by definition we can set $\tilde {h}_H=0$ ).
Now since $W(H)$ is trivial, from (1-6) and (1-7) we have:
$$I_{\lambda }^{\Lambda }=\cases{\epsilon (\omega ),&if
$\lambda=\omega (\Lambda +{\bf \rho }_G)$;\cr 0 &otherwise.\cr}
\eqno (2-2)$$

$\Lambda $will always appear in thmbination $\hat {\Lambda }=\Lambda +\rho_G$
and for simplicity drop
hat  on it . So (2-2) can be written in the compact form:
$$I_{\lambda }^{\Lambda }=\sum _{\omega \in W(G)} \epsilon (\omega )
\delta_{\lambda }^{\omega (\Lambda )}\eqno (2-3)$$
If we insert (2-3) into (1-5), it follows that:
$$M^{\Lambda,\Lambda '}=\sum _{\omega ,\omega '\in W(G)}
\epsilon (\omega )\epsilon (\omega ')P^{\omega (\Lambda ),\omega '(\Lambda
')}\eqno (2-4)$$
This relation gives the main tool to obtain a modular invariant
$M^{\Lambda ,\Lambda '}$ for $\hat G$  through a modular invariant
$P^{\lambda ,\lambda '}$ of the free bosonic theory.  Partition functions of
the $\hat H$-theory are of the form:
$$ {\cal Z}_{\cal L}(\tau ,{\bar \tau } )=[\eta (\tau ){\bar \eta } ({\bar \tau
} )]^{-l}
\sum_{({\bf p}_L,{\bf p}_R)\in {\cal L}}q^{{1\over 2}{\bf p}_{L}^{2}}{\bar
q}^{{1\over 2}{\bf p}_{R}^{2}}
\eqno (2-5)$$
where $\cal L$ is the Lorentzian even self-dual lattice of dimension
$(l,l)$:
$$ {\cal L}=\Bigg \{ ({\lambda \over {\sqrt g}}+\sqrt g{\bf v},
{\lambda \over {\sqrt g}}+\sqrt g{\bf v}');\ {\bf v,v'}\in \Gamma (G),\ \lambda
\in
{\Gamma ^\ast (G)\over {g\Gamma (G)}}\ \Bigg \} .\eqno (2-6)$$
$ {\cal Z}_{\cal L}(\tau ,{\bar \tau })$ can be considered as the diagonal
combination
of left and right characters of $\hat H$, which (for the left sector) is
defined
to be:
$$\chi^{(H)}_{\lambda }(\tau )=[\eta (\tau )]^{-l}\sum _{{\bf v}\in \Gamma
(G)} q^{{1\over 2}({\lambda \over {\sqrt g}}+\sqrt g{\bf v})^2}\eqno (2-7)$$
We call $\cal L$ the diagonal lattice. From (2-7) it is obvious that
$\chi^H_{\lambda }(\tau )$
is invariant under translation of $\lambda $ in the lattice $g\Gamma (G) $,
so when dealing with $\hat H$-theory, we can consider each weight $\lambda $
modulo $g\Gamma (G) $.

 Consider a Lorentzian even self-dual lattice ${\cal L}'$ which  includes
$\sqrt (g\Gamma,g\Gamma (G))$ as a sublattice, then
$ {\cal Z}_{{\cal L}'}(\tau ,{\bar \tau })$ is a  modular
invariant
combination of left and right characters, specified  by
$P^{\lambda ,\lambda '}$. We can construct such ${\cal L}'$ lattices from a
given
lattice $\cal L$ (e.g. diagonal lattice) by using an $(l,l)$
dimensional vector $ \sigma =(\sigma _L,\sigma _R) $
(called shift vector), with the following properties:
$$\left. \eqalign {a)\quad &\sigma \in ({ \Gamma ^\ast (G)\over {\sqrt g}},
{ \Gamma^\ast (G)\over {\sqrt g}})\cr
b)\quad &\exists n \in Z\  {\rm such  \  that:}\ n \sigma
\in {\cal L}\cr
c)\quad &\sigma .\sigma \in 2Z \cr}\right. \eqno (2-8)$$

If $n $ is the smallest possible integer with property
(2-8b), then we can see that the lattice
$${\cal L}'=\cup_{j=0}^{n -1}\ ({\cal L}_0+\jmath \sigma ) \eqno (2-9)$$
would have the required properties, provided
$${\cal L}_0=\{ {\bf v}\in {\cal L}\ ;\ {\bf v}.\sigma \in Z\} \eqno (2-10)$$

In order to perform this procedure, we can
assume $\sigma $ to have the form:
$$\sigma =({{\bf a} \over {\sqrt g}}\ ,\ {{\bf a+b}\over {\sqrt g}})\eqno
(2-11)$$
where $\bf a$ and $\bf b  $ are two arbitrary weights with the constraint
(2-8c) taking the form :
$${\bf b^2+2a.b}=0\pmod {2g}\eqno (2-12)$$
resulted from (2-8c). define n to be the smallest integer with the property:
$$n {\bf b}\in g\Gamma (G)\eqno (2-13)$$
and  $q$  the smallest integer such that $q{\bf b} \in \Gamma $
\footnote*{In this paper by "root" we mean elements of "root lattice".}
(For example in $ SU(N) $, $q$ is some divisor of $N$). Then it is clear that
$q  $ divides $n$.
Suppose $p$ is the greatest divisor of $g$ with the property
$$q{\bf b}=p\beta
 \eqno (2-14) $$
 where $\beta$ is a root. $p$ is obviously prime relative to $q$ and we have
$$n ={qg\over p}\eqno (2-15) $$

 To find the lattice ${\cal L}'$ (and hence
$P^ {\lambda ,\lamb}$), we should first find the sublattice ${\cal L}_0$.
$({\bf v}.\sigma )$ being an integer implies that
Each ${\bf v}\in {\cal L}_0$ has of the form:
$${\bf v}=(\lambda_0/\sqrt g\ +\sqrt g{\bf u}\ ,\
\lambda_0/\sqrt g\ +\sqrt g{\bf u'})\eqno (2-16a) $$
such that:
$$\lambda_0.{\bf b}=0\pmod g\eqno (2-16b) $$

 From (2-9) we have:
$${\cal L}'=\Bigg \{ ({\lambda_0+j\sigma_L\over {\sqrt g}}+\sqrt g{\bf u},
{\lambda_0+j\sigma_R\over {\sqrt g}}+\sqrt g{\bf u'});\ {\bf u,u'}\in \Gamma
(G),
\ \ \ \ \ \ \ \ \ \ \ \ \ \ \ \ \ \ \ \ \ \ \ \ \ \ \ \ \  $$
$$\ \ \ \ \ \ \ \ \ \ \ \ \ \ \ \ \ \ \ 0\leq j\leq n -1, \ \lambda_0\in
{\Gamma^\ast (G)\over {g\Gamma (G)}},  \ \ \lambda_0.{\bf b}=0\pmod g \Bigg
\}\eqno (2-17)$$

Using (2-5) and (2-7) we see that
$Z_{\cal L'}(\tau,{\bar \tau})$ has the form of (1-3) with

$$P^{\lambda ,\lambda '}=\sum _{\lambda_0\in {\Gamma^\ast (G)\over {g\Gamma
(G)}}}\ \ \sum _{j=0}^{n -1}\delta ^{\lambda ,\lambda_0+j\sigma_L}
\delta^{\lambda ',\lambda_0+j\sigma_R}\delta_{(g)}^{\lambda_0.{\bf b},0}\eqno
(2-18) $$
where  $\delta _{(g)}^{a,b } $ we impose the equality of indices a and b modulo
$g$. Since
$\lambda ,\lambda '$ and $\lambda_0$ all belong to the same set,
summing over $\lambda_0$ and using
(2-11) and (2-11) gives:
$$P^{\lambda ,\lambda '}=
  \sum _{j=0}^{n -1}\delta^{\lambda ',\lambda +j{\bf b}}
\delta_{(g)}^{\lambda .{\bf b +b^2/2},0}\eqno (2-19)$$

To write (2-19) in a more compact          form, we introduce the set of
weights
$$\bar \lambda_{\imath }=\lambda +j_{\imath }{\bf b}\eqno (2-20)$$
where $j_{\i }$'s are integer solutions to the equation
$$(\lambda +{1\over 2}j_{\imath }{\bf b}).{\bf b}=0\pmod g\eqno (2-21) $$
in the range  $[0,n-1]$. We call the set of $\lb_{\i }$'s {\it weights
associated
to }$\lambda$. Then (2-19) simplifies to
$$P^{\lambda ,\lambda '}=\sum _{\imath}\delta^{\lambda ',\bar \lambda_{\imat
\eqno (2-22)$$

Before proceeding further we make two observations;  First, from (2-22) we see
that for a given pair $(\lambda
,\lambda')$,
$\pl $ is either zero or one and  second, from (2-20) and (2-21) we see that
$\pl $ is independent of the choice of $\bf a$ in (2-11),and
and hence all subsequent results can be written in terms of the weight $\bf b$
only.
So hereafter by shift vector we mean the vector $\bf b$.

In order to solve (2-21) for $j$, we use (2-14) and (2-15) to get the integer
equation
$$\lambda .\beta +j_{\imath }t=0\pmod n \eqno (2-23) $$
where
$$t={1\over 2}{\bf b}.\beta \eqno (2-24) $$
As a consequence   of (2-12) and (2-14a) t is an integer which  should be
imposed
on shift vector $\bf b $ from the beginning. Now let
$$\mu =[t,n ]\eqno (2-25) $$
(By $[\ ,\ \cdots ]$ we denote greatest common divisor of two or more
integers.)
Equation (2-23) has integer solution for $j$ if
$$\mu \vert \lambda .\beta \eqno (2-26) $$
Weights not satisfying (2-26) are absent from modular invariant. We call
weights
which do satisfy (2-26) as {\it allowed weights}.
Assuming $\lambda $
to be allowed, we can divide both sides of (2-22) by $\mu$. the resulting
equation
has the unique solution:
$$j_0=-t'^{-1}({\lambda .\beta \over {\mu }})\pmod {n '}\eqno (2-27) $$
in the range $[0\ ,\ n '-1] $, where $t'=t/\mu $ and $n '=n/\mu $. The
integer
$t'^{-1}$ is
the inverse of $t'$ in the group ${ Z^*}_{n'}$ defined to be the set of
integers prime to $n '$
modulo $n'$. The set of all $j_{\imath }$'s
in the range  $[0,n-1]$ are:
$$j_{\imath }=j_0+\imath n '\ ;\qquad
\imath =0,\cdots \mu -1\eqno (2-28) $$
and the set of associated weights to (allowed) $\lambda $ are:
$$\bar \lambda_{\imath }=\bar \lambda_0+\imath n '{\bf b}\ ; \qquad
\imath =0,\cdots \mu -1 \eqno (2-29) $$
where
$${\bar \lambda}_0=\lambda -t'^{-1}({\lambda .\beta \over {\mu }}){\bf
b}\eqno30) $$
Since $\lambda .\beta $ is a multiple of $\mu $, associated weights to an
allowed
$\lambda $ are also allowed weights.

In the special case where $n $ and $t$ are relatively prime,
we have $\mu =1$, and all weights $\lambda $ are allowed. Moreover for each
$\lambda $ there is one and only one associated weight $\bar \lambda $.

Due to linearity of (2-30) with respect to $\lambda $, associated weight
to $\sum_{\jmath }m_{\jmath }\gamma_{\jmath }$, where $\gamma_{\jmath }$'s
are allowed weights, is
$$\overline{(\sum_{\jmath }m_{\jmath }\gamma_{\jmath })}=
\sum_{\jmath }m_{\jmath } \overline{(\gamma_{\jmath })} \eqno (2-31) $$
where the  bar on a weight means the weights associated  to it (in fact
on of the set of the associated weights).
So there exists a linear operator $T$ which acting on an allowed weight, gives
one of the  weights associated to it:
$$\bar \lambda_{\imath }=T\lambda +\imath n '{\bf b}\ ;\qquad
\imath =0,\cdots \mu -1\eqno (2-32) $$

Since $n {\bf b}\in g\Gamma (G) $, shiftings (of an arbitrary weight) by
multiples of $ {\bf b}$ makes a cyclic group of order $\mu$. The action of
the generator of the group on a weight is represented by a nonlinear operator
$\xi $ such that
$$\xi \lambda =\lambda + n'{\bf b}\ .\eqno (2-33)$$
Hence, (2-32) can be written in the form:
$$\bar \lambda_{\imath }=\xi^{\imath }T(\lambda )\ ;\qquad
\imath =0,\cdots \mu -1\eqno (2-34) $$
We shall call the set of weights obtained by repetetive action of $\xi $ on an
allowed weight $\lambda $, as $\lambda $-{\it family}.
Each $ \lambda $-family consists of $\mu $ elements.

Now in order to exhibit the operator $T$ explicitly, we represent weights by
$ket$'s
 (column vectors) and roots by $bra$'s (row vectors), with the conventional
inner product between them
(Recall that roots are vectors in the co-root lattice ,  and the co-root
lattice is
the dof the weight lattice).
 Then, by using (2-14) and (2-30) we have:
$$T=1-{{t'}^{-1}p\over {\mu q}}\vert \beta > < \beta \vert \eqno (2-35) $$

If we decompose weights in the basis of fundamental weights denoted by
$\alpha ^{\imath} $'s, and roots in the basis of simple
roots denoted by $\alpha _{\imath} $'s (such that $\alpha_{\i }.\alpha^{\j }
=\delta_{\i }{}^{\j })$, elements of $T$ are given by:
$$T_{\imath }{}^{\jmath }=\delta_{\imath }{}^{\jmath }-
{t'^{-1}p\over {\mu q}}r_{\imath }r^{\jmath }\eqno (2-36) $$
where $r_{\imath } $'s and $r^{\jmath }$'s are components of $\beta $  in the
basis of fundamental weights and
simple roots, respectively . The operator $T$ is manifestly a self-adjoint
operator
and can act on weights (column vectors) from the right or on roots (row
vectors) from the left. It should be also noted that $T$ acts only on allowed
weights.

Now let us further clarify the structure  of the $\hat H $ modular invariant
partition functions. From (2-32) one sees that all elements of $\bar {\lambda}
$
are coupled to $\lambda $. It is also true that all members of $\lambda
$-family are
coupled to $\bar {\lambda }$. To prove this point we can either use a property
of T
as an operator on the families or use the symmetry of $P^{\lambda,\bar \lambda}
$ .
To see the symmetry of P let us write the  equations for $\lambda $and its
associated weight as following:

$$\left\{ \eqalign {\bar \lambda -\lambda &=0\pmod {\bf b}\cr (\bar \lambda
-\lambda).
\beta &=0\pmod {2n }\cr}\right. \eqno (2-37) $$
It is easy to show that each $\bar \lambda $ satisfying (2-37) belongs to the
set of $\bar \lambda_{\imath }$'s ,  and vice versa.

(2-37)  can be written in the equivalent following form

$$P^{\lambda ,\lambda '}=\delta_{\bf b}^{(\bar \lambda -\lambda),0}
\delta_{2n }^{(\bar \lambda +\lambda ).\beta ,0}\eqno (2-38)$$
In this form, symmetry bet left and right weights in a modular
invariant is manifest; that is:
$$P^{\lambda ,\lambda '}=P^{\lambda ',\lambda }\eqno (2-39) $$
  Now we are able to state our theorem on the structure of the $\hat H $
modular invariant partition functions.

{\bf Theorem 2-1)}
{\it $\hat H $ modular invariant partition functions consist of products of
left and right characters over mutually associated families which are obtained
from each other by the action of T }.

{\bf Proof:} We have seen that all members of  ${\bar \lambda} $
family are associated to $\lambda $. Furthermore..  the operator T which
generates weights associated to $\lambda $ generates elements of $\lambda
$-family when applied to any element of ${\bar \lambda} $ family.
We see  from (2-35) that $\bf h$ is an eigenvector  of $T$ with integer
eigenvalue. So, $T$ acting on $\xi \lambda =\lambda +n '{\bf {b}} $ gives a
member
of the associated family
to $\lambda $, that is a member of the set  $\{ \xi^{\imath }T(\lambda
)\}_{\imath =0}^{a-1}$.
Therefore the same family is associated to all members of $\lambda $-family.
This shows that the whole $\lb $-family is associated to the whole $\lambda
$-family.
that is
$${\cal Z}(\tau )=\sum _{\lambda \ allowed }\sum _{\bar \lambda \ associated
 \       to \lambda}
\chi _{\lambda }(\tau )
{\bar \chi}_{\bar \lambda '}(\bar \tau )=\sum_{\lambda :\mu \vert \lambda
.\beta }
F_{\lambda }(\tau )\bar F_{T(\lambda )}(\bar \tau )  \eqno (2-40) $$
in which $F_\lambda $ is the character over the space of $\lambda $-family

$$ F_{\lambda }(\tau )=\sum_{\imath =0}^{\mu -1}\chi_{\xi^{\imath }(\lambda )}
(\tau )\eqno (2-41) $$
This completes the proof.

Allowed families are pairwise associated to each other. That is because $T^2$
is effectively equivalent to unity when it is viewed as acting on the families.
To see this, we can apply T twice on a $ \lambda $ btain

$$T^2\lambda =\lambda +2t'^{-1}\lbm (t'^{-1}t'-1){\bf b}.$$

Noting that  $t'^{-1}$ is the inverse of t in the group $ {Z^\ast }_{n'} $ i.e
, $(t'^{-1}t'-1)$ is a multiple of
$n'$. and  using (2-26) we deduce that $T^2\lambda $ belongs to $\lambda
$-family,
which means that $T^2 \equiv 1$.

The second theorem which would be useful for some future applications helps us
to remove the   degeneracy                   of the solutions ,
identifying some repeated modular invariants.

{\bf Theorem 2-2)} {\it Each multiple $r{\bf b}$ of a shift vector ${\bf b}$
would
yield the same
modular invariant, provided that:}
$$[r,qg]=1\eqno (2-42) $$.
(Due to its detailed and ltechnical proof we defer the it to  the appendix A.)

The condition (2-42) on $r$ requires that $q$, $p$ and $n $ are the same for
both
modular invariants, and under  ${\bf b}\to r{\bf b} $ we have only $\beta \to
r\beta $.
So in order to avoid overcounting, we should choose $\bf b$ such that the
root $\beta $ would not be a multiple of some other root.
In other words components of $\beta $ in the basis of simple roots should
be relatively prime.

Now we are prepared to obtain the modular invariant partition functions
of $\hat G$ from $\hat H$-theory. This is formally represented in (2-4),but
in practice in the process of its implementation one confronts serious
difficulties .
First, notice that the summation over Weyl group of $G$ makes the number of
terms in a
modular invariant of $\hat G$ too much bigger than that of $\hat H$.
Second, it is not easy to find a simple relation like (2-34) that gives
the  mutually associated weights.
Finally, minus signs appear in modular invariants of $\hat G$ because of
$\epsilon (\omega ) $ and $\epsilon (\omega ') $ in (2-4), which
make it unphysical. In order to have a physically acceptable partition
function, we
should combine such modular invnts to cancel out unwanted negative terms.
In general it is not possible to do this; and moreover, there is not a	simple
way to
distinguish cases for which we can really construct a partition function.

However, there are special cases which we can overcome all these difficulties
and
derive modular invariant partition functions. In the next two sections we will
give
examples where this cancellations occurs.

\vskip 0.5cm

{\bf{3-\ \ $D$-series of $SU(N)_k$}}

\vskip 0.5cm
The D-series of $SU(N)$ is obtained by a particular choice of the shift vector.
Before introducing these shift vectors we first review some special features of
$SU(N)$ which is needed in our discussion.

$SU(N)$ is a simply- laced group and therefore its co-root lattice is
 the same as root lattice. Simple roots $\{ \alpha_{\imath } \}_{\imath
=1}^{N-1}$
are the basis for the  root lattice, and in the conventional normalization have
inner
products:
$$g_{\imath \jmath }=\alpha_{\imath }.\alpha_{\jmath }=
\cases {2,&if$\ \ \vert \imath -\jmath \vert =0$;\cr
-1,&if $\ \ \vert \imath -\jmath \vert =1$;\cr 0,&otherwise\cr}\eqno (3-1) $$

Weight lattice has the basis $\{ \alpha^{\imath } \}_{\imath =1}^{N-1}$  where
$ \alpha^{\imath } $'s
are fundamental weights, with inner products:
$$g^{\imath \jmath }=\alpha^{\imath }.\alpha^{\jmath }=\{ inf(\imath ,\jmath )
-{\imath \jmath \over {N}}\}\eqno (3-2) $$.

We use $g^{\imath \jmath }$ and its inverse $g_{\imath \jmath }$
to raise and lower indices. Note that although the indices on $\alpha $ 's
are not component indices we still have ,
 $$ \alpha^{\imath }=g^{\imath \jmath }\alpha_{\jmath }\eqno (3-3) $$.
Weyl group of $SU(N)$ is a finite group of order$ N !$, with $l=N -1 $
generators,
which are the Weyl reflections with respect to simple roots.
 Weyl reflection with respect to the simple  root
$\alpha_{\imath }$
acts on the weight_1,m_2,\cdots ,m_l) $ as follows:
$$\omega_{\alpha_{\imath }}(m_1,m_2,\cdots , m_l)=
(m_1,\cdots ,m_{\imath -1}+m_{\imath },-m_{\imath },m_{\imath +1}+m_{\imath }
,\cdots ,m_l)\eqno (3-4) $$
where $m_i$ 's are the Dynkin indices .

The $D$-series are obtained taking the shift vector in the lattice
$g\Gamma ^{\ast}(G)$; that is ${\bf b}=g\phi$, where
$\phi$ is any weight. Noting that $\rm b$ is defined modulo $g\Gamma (G)$
we can restrict $\phi$ to be in the cell $\Gamma^{\ast }(G)/\Gamma (G)$.
Hence we can choose $\phi$ to be $g\alpha^{\nu}\ \ (\forall \nu =1,\cdots l)$.
Using (3-2)and (3-3) we see that $\alpha^{\nu }-\nu \alpha^1 \in \Gamma(G)$.
Therefore independent choices for $\rm b$ in the lattice $g\Gamma^{\ast } (G)$
are of the form:
$${\bf b}=g\nu \alpha^1 \eqno (3-5)$$

We should follow the procedure of the previous section to find integers
$q,p,n;t$ and the vector $\beta$. Let $\nu_0$ be g.c.d. of $N$ and $\nu$,
such that $N=\nu_0N'$ and $\nu =\nu_0\nu'$; and $N'_0$ be the g.c.d. of
$N'$ and $g$. Then from the definitions given in section 2 it turns out
that $q=N'/N'_0$ and $p=g/N'_0$. Using (2-14) and (2-15)
we obtain $n=N'$ and $\beta =\nu 'N\alpha^1$. If we choose
${\rm b}=g\nu_0\alpha^1$ instead of (3-5), the only difference would be
in $\beta$ which whould be $N\alpha^1$. From theorem (2-2) the same modular
invariant would emerge in both cases.
So we assume $\nu$ to be one of the divisors of N from the beginning
  , for example if $N$ is a prime we only have the choice $\nu =1$.
Thus the only characteristic of the modular invariant (which we shall see to
lead to the $D$-series) is the number $\nu$ which is a   divisor             of
$N$.
The other auxiliary quantities are:
$$\left\{ \eqalign{{\bf b}&=\nu g\alpha^1\cr n &=N'\cr \beta &=N\alpha^1\cr
t&=\nu g(N-1)/2\cr}\right. \qquad ;N=\nu N'\eqno (3-6) $$

Then the resug associated weights to
$\lambda =(m_1,m_2,\cdots ,m_l) $ are given by:
$${\bar \lambda} =\lambda +j\nu g\alpha^1\eqno (3-7) $$
where $j$ is any of the integer solutions of the equation
 $$ \lambda .\beta +jt=0\pmod {N'}\eqno (3-8) $$

 From (3-3) we have: $\beta .\lambda =\sum_{\imath }(N-\imath )m_{\imath }$.
But in (3-8) we need only to know $(\beta .\lambda \bmod {N'})$.
So instead of (3-8) we can solve ,
$$N(\lambda )-jt=0\pmod {N'}\eqno (3-9) $$
where the quantity
$$N(\lambda )=\sum_{\imath =1}^l\imath m_{\imath }\eqno (3-10) $$
is called $N$-$ality$ of the weight $\lambda$.

Equation  (3-7) can also be written in the form
$${\bar \lambda }=\eta^{j\nu }(\lambda )\eqno (3-11) $$
Where the (nonlinear) operator $\eta $ is defined such that:
$$\eta (m_1,m_2,\cdots ,m_l)=(m_1+g,m_2,\cdots ,m_l)\eqno (3-12) $$
Obviously $\eta $ is the generator of a $Z_N$ group and $\eta^{\nu }$
also generates a $Z_{N'}$ group.

The whole modular invariant in $\hat H$-Theory given by (3-6) is:
$$D^H(\tau ,{\bar \tau })=\sum_{\scriptstyle\lambda ,j\atop\scriptstyle
 N(\lambda )=jt\pmod {N'}}
\chi_{\lambda }(\tau ){\bar \chi }_{\eta^{j\nu }(\lambda )}(\bar \tau )\eqno
(3-13) $$

We should carry out this modular invariant to $\hat G$-theory via relation
(2-4).
To do this, we should find associated weights to each allowed $\lambda =\omega
(\Lambda )$,
and then for each $\bar \lambda $ find appropriate $\omega '$
such that ${\bar \lambda }=\omega '(\Lambda ') $ where $\Lambda '$ is some
weight in the
fundamental domain of affine $G$ at height $g$ (denoted by $B_g$). Such an
$\omega '$ always exists and
is {unique.}\footnote*
{It should be noted that the cell ${\Gamma ^\ast (G)\over {g\Gamma (G)}}$
consists of $B_g$
together with all its Weyl reflections [6]. So for every weight in this cell,
there is one and only one Weyl reflection that can fold it back  on the
fundtal
domain.}
So (in principle) we should repeat the procedure of finding associated
weights in $H$-theory for all allowed Weyl transformations of $\Lambda $.
But here we have a simplifying property, according to which it is enough
to find associated weights only for $\Lambda $ itself.
To the see this, consider a simple Weyl reflection of an arbitrary weight
$\lambda $.
We see from (3-4) and (3-10) that
$$N(\omega_{\alpha_{\imath }}(\lambda ))=N(\lambda )\eqno (3-14) $$
So in (3-9) the same set of $ j $'s would emerge
for both $\lambda $ and $\omega_{\alpha_{\imath }}(\lambda )$, or symbolically:
$$j_{\omega_{\alpha_{\imath }}(\lambda )}=j_{\lambda } $$
 From (3-7) we deduce:
$$\overline {\omega_{\alpha_{\imath }}(\lambda )}=\omega_{\alpha_{\imath
}}(\bar
\lambda )
\pmod {g\Gamma (G)}\eqno (3-15) $$
Since elements of Weyl group can be identified as products of simple
Weyl reflections, (3-15) can be generalized to:
$$\overline {\omega (\lambda )}=\omega (\bar \lambda ),\eqno (3-16) $$
where $\omega $ is any element of $W(G)$, and we have dropped the trivial
modding.
 Property (3-16) can also be written in the form:
$$P^{\omega (\lambda ),\omega (\lambda ')}=P^{\lambda ,\lambda '}\eqno (3-17)
$$
We call property (3-17) as the invariance of modular invariant $P^{\lambda
,\lambda '}$
under Weyl group of $G$.

 From (3-17) we see that for each weight of the form $\lambda =\omega (\Lambda
)$
the same set
of associated weights appears. So $\vert W(G)\vert $ equal terms
in modular invariant of $\hat G$ would be resulted. In other words, (3-17)
together with (2-4) gives:
$$M^{\Lambda ,\Lambda '}=\vert W(G)\vert \sum_{\omega '\in W(G)}\ \ \epsilon
(\omega ')P^{\Lambda ,\omega '(\Lambda ')}$$
Dropping the overall factor $\vert W(G)\vert $, and using (2-17), we have:
$$M^{\Lambda ,\Lambda '}= \sum_{\omega ,\imath } \epsilon
(\omega )\delta ^{\omega ({\bar \La }_{\imath }),\Lambda '}$$
But, as noted before, in this summation only one $\omega $ which we call
it $\omega^{\Lambda ,\imath }$ can fold ${\bar \Lambda }_{\imath }$ back onto
the fundamental domain of $G$. Therefore we have :
$$M^{\Lambda ,\Lambda '}= \sum_{\imath } \epsilon (\omega^{\Lambda ,\imath })
\delta ^{{\tilde \Lambda }_{\imath },\Lambda '}\eqno (3-18)$$
where ${\tilde \Lambda }_{\imath }$ is some weight in the fundamental domain
$B_g$, such that:
$${\tilde \Lambda }_{\imath }=\omega^{\Lambda ,\imath }({\bar \Lambda }_{\imath
})
\eqno (3-19) $$

We see that the idea of associated weights can still be used for modular
invariants of $\hat G$-theory. The only difference is that, in this case we
should perform one more step to find $G$-associated weights from
$H$-associated
weights via relation (3-19), followed by imposing appropriate sign to the
resulted
term in modular invariant due to {$\epsilon (\omega^{\Lambda ,\imath })$.}
\footnote* {This feature, which lead to (3-21) is not special for the shift
vector
considered in this section. There also can be found some other examples
with this property.}

Consider $\hat H$-associated weights to $\Lambda $ in $B_g$ resulting from
(3-9) and
(3-11):
$$\left\{ \eqalign{N(\Lambda )-jt&=0\pmod {N'}\cr
{\bar \Lambda }_j&=\eta^{j\nu }(\Lambda )\cr}\right.\eqno (3-20) $$
In order to find $\tilde \Lambda $'s from $\bar \Lambda $'s, we use the
following
element of Weyl Group:
$$\omega_t=\omega_{\alpha_l}\omega_{\alpha_{l-1}}\cdots \omega_{\alpha_1}\eqno
(3-21) $$
whose action on a weight is as follows:
$$\omega_t(m_1,m_2,\cdots m_l)=(m_2,\cdots m_l,-\sum_{\imath =1}^lm_{\imath
})\eqno (3-22)$$
We also define the following operator:
$$\sigma (m_1,m_2,\cdots m_l)=(g-\sum_{\imath =1}^lm_{\imath },m_1,\cdots
m_{l-1}),\eqno (3-23) $$
which is the generator of the $Z_N$ subgroup  of the outer automorphism the
Kac-Moody algebra  that transforms weights belonging
to an $N$-member family in $B_g$ into each other.
It is easy to see that $\omega_t$ relates operators $\sigma $ and $\eta $
(as generators of described $Z_N$-groups in $\hat G$ and $\hat H$
theories , respectively)
in the following  way:
$$\eta =\sigma \omega_t=\omega_t \sigma \pmod{g\Gamma (G)}\ . $$
So for each power of $\eta $ we have:
$$\eta^m=\omega_t^m \sigma^m.\eqno (3-24) $$
 From (3-24) it is obvious that:
$$\epsilon (\omega_t)={(-1)}^l.\eqno (3-25) $$
Clearly we have $\omega _t^{-j\nu }$ belonging to $W(G)$ is the desired element
$\omega ^{\Lambda ,i}$since when it acts on $\Lambda $ gives $\sigma ^{jl}
(\Lambda)$ which belongs to $B_g$ therefore $\tilde \Lambda _j=\sigma ^{j\nu}
(\Lambda) $and (3-20) can be replaced by
$$\left\{ \eqalign{N(\Lambda )-jt&=0\pmod {N'}\cr
{\tilde \Lambda }_j&=\sigma^{j\nu }(\Lambda )\cr}\right. \eqno (3-26) $$
The corresponding term in the modular invariant has the sign
$$\epsilon (\omega^{\Lambda ,\imath })={(-1)}^{lj\nu } =(-1)^{(N-1)j\nu} .\eqno
(3-27) $$
Therefor the final modular invariant in $G$-theory is:
$$M(\tau ,{\bar \tau })=\sum_{\scriptstyle \Lambda \in B_g;0\leq j<N'
\atop\scriptstyle N(\Lambda )=jt\pmod {N'}}
{(-1)}^{lj\nu }\chi_{\Lambda }(\tau ){\bar \chi }_{\sigma^{j\nu }
(\Lambda )}(\bar \tau ).\eqno (3-28) $$
Notice that at this stage many of the intermediate forms have gone out of the
way and the modular invariant has acquired a simple form.
This is not in general a partition function.
For odd $N$ however, $l$ is even and no minus sign appears. Therefore (3-28)
for odd $N$ is really a partition function as follows:
$$D_{\nu }(\tau ,{\bar ,\tau })=\sum_{\scriptstyle \Lambda \in B_g;0\leq j<N'
\atop\scriptstyle N(\Lambda )=jt\pmod N'}
\chi_{\Lambda }(\tau ){\bar \chi }_{\sigma^{j\nu }(\Lambda )}(\bar \tau ).\eqno
(3-29) $$

We recthat $\nu $ is any divisor of $N$ (taking  into account the trivial
divisor $\nu =1$,
but not $n=N$ which obviously leads to $A$-series).
It is also noticeable that oddness of $N$ guarantees integrality of
$t={1\over 2}\nu g(N-1)$ for any choice of $\nu $ and $g$.

For even $N$, we should  eliminate minus signs in (3-28). The first
thing that we can do is to choose $\nu $ even. This choice at one hand makes
$t$
an integer , and on the other hand drops all negative terms in (3-28),
giving again a partition function in the form of (3-29) but restricting $\nu $
to even  divisors of $N$. Due to t being an integer
For odd $g$ this is the only way to have a $D$-partition function.
For even $g$, however we have another chance . In this case evenness of
$g$ makes $t$ integer, and we are allowed to choose $\nu $ an odd divisor
of $N$. But now minus signs do appear in (3-28). What we can do, is to subtract
this modular invariant from $D_{2\nu }$-partition function,
for which $t$ is
twice and  $N'$ is half of the first one.
Fortunately in the resulting modular invariant all negative terms cancel
out,  and we again obtain   a partition function ,
$$D'_{\nu }(\tau ,{\bar ,\tau })=
\sum_{\scriptstyle \Lambda \in B_g;j{\rm odd}
\atop\scriptstyle N(\Lambda )=jt\pmod {N'}}
\chi_{\Lambda }(\tau ){\bar \chi }_{\sigma^{j\nu }(\Lambda )}(\bar \tau )\ \ \
\
\ \ \ \ \ $$
$$\ \ \ \ \ \ \ \ \ \ \ \ \ \ \ \ \ \ \ \ \ \ +\sum_{\scriptstyle \Lambda \in
B_g;j{\rm even}
\atop\scriptstyle N(\Lambda )-jt=N'/2\pmod {N'}}
\chi_{\Lambda }(\tau )
{\bar \chi }_{\sigma^{j\nu }(\Lambda )}(\bar \tau ).\eqno (3-30) $$

As a final note in this section let's consider the case of $SU(4)$ as an
example. For arbitrary $g$, only one partition function of the form (3-29)
would
be  resulted from  the
choice $n=2$,  which leads to $t=3g$ and $N'=2$. So we have:
$$D_2(\tau ,{\bar \tau })=\sum_{ambda )=0}
\vert \chi_{\Lambda }(\tau )\vert^2+\sum_{N(\Lambda )=g\bmod 2}
\chi_{\Lambda }(\tau ){\bar \chi }_{\sigma^2(\Lambda )}(\bar \tau ).$$

For even $g$, there is another partition function of the form (3-30)
resulted from the choice $\nu =1$.
Supposing $g=2g_0$ we have $t=3g_0$ and $N'=4$. The resulted
partition function is:
$$D'_1(\tau ,{\bar \tau })=
\sum_{\scriptstyle j=1,3\atop\scriptstyle N(\Lambda )=-jg_0\bmod 4}
\chi_{\Lambda }(\tau ){\bar \chi }_{\sigma^j(\Lambda )}(\bar \tau )
\sum_{\scriptstyle j=0,2\atop\scriptstyle N(\Lambda )+jg_0=2\pmod 4}
\chi_{\Lambda }(\tau ){\bar \chi }_{\sigma^j(\Lambda )}(\bar \tau ) $$

\vskip 10pt

{\bf {4- \  A-D-E Classification for $SU(2)$ :}}

\vskip 5pt

In this section we want to use  our method for affine $SU(2)$  at arbitrary
level $k$ (or height $g=k+2$). The special feature of $SU(2)$ is that, it
is a rank-1 group with a one dimensional root lattice. Consequently,
we can show that in this case mono-shifts are enough to produce all
possible Lorentzian self-dual lattices. For this purpose we have shown
in appendix-B that each arbitrary  Lorentzian even self-dual lattice
${\cal L}'$ yields the same result which we will obtain by the mono-shift
method.

Before deriving the general form of  modular invariants, let's look at
$D$-partition
function of $SU(2)$. For $N=2$, only divisor $\nu =1$ can be chosen.
So only for even $g$'s $D$-partition function of type (3-30) exists. From (3-9)
we have $N'=2$ and $t=g_0$ where $g=2g_0$. Now (3-30) gives:
$$D'_1(\tau ,{\bar \tau })=\sum_{m=g_0\pmod 2}
 \ \ \chi_m(\tau ){\bar \chi }_{g-m}(\bar \tau )+\sum_{m=1\pmod 2}
\ \vert \chi_m(\tau )\vert^2 \eqno (4-1) $$
where each weight $\Lambda $ is specified by its only component
$m\  (\Lambda =m\alpha^1)$, and we have used:
$n(\Lambda )=m$ and $\sigma (\Lambda )=(g-m)\alpha^1$. We can break (4-1)
into two follo cases:
$$D_{g=2g_0}(\tau ,{\bar \tau })=\cases {\sum_{m\ {\rm even}}
\chi_m(\tau ){\bar \chi }_{g-m}(\bar \tau )+\sum_{m\ {\rm odd}}
\vert \chi_m(\tau )\vert^2 &$g_0$ even\cr
\sum_{m\ {\rm odd}}\vert \chi_m(\tau )+\chi_{g-m}(\tau )\vert^2
&$g_0$ odd\cr}\eqno (4-2) $$

Now in order to find all modular invariants of $SU(2)$ at a given height $g$,
we should consider all shift vectors $h={p\over q}\beta $. But according to
the
note given
after theorem (2-2), $\beta $ should be a root which is not a multiple of
some other root. So the only choice for $\beta $ is $\beta=\alpha_1$.

For choosing $q$, we note that in $SU(2)$ at most 2 times a weight will
be a root. So $q$ can be 1 or 2. But a little calculation shows that $q=2$
does not give integer $t$. Therefore only $q=1$ is acceptable.

Fixing choices for $q$ and $\beta $, each  modular invariant is characterized
only by
$p$ which is a divisor of $g$. Related quantities in terms of $p$ are:
$$\left\{ \eqalign{{\bf b}&=2p\alpha^1\cr n &=g/p\cr t&=p\cr}\right. \eqno
(4-3)
$$
Supposing $p=p'\mu $ and $n =n '\mu $ where  $\mu =[p,n]$,
$T$-matrix is just the number:
$$T=1-2{p'}^{-1}p'\eqno (4-4) $$
where ${p'}^{-1}$ is the inverse of $p'$ in group $Z^{\ast }_{n'}$.
In other words, we can find associated weight (or family) to a weight
just by multiplication with the number (4-4).

The lattice $g\Gamma (G)$ is the set of multiples of $ g\alpha_1 $.
So in the basis of $\alpha^1$, weights differing by $2g$ are identical.

Now using (2-25) we can write each arbitrary modular invariant $P^{m,m'}$
(standing for $ P^{\lambda ,\lambda '}$)
in the form:
$$P^{m,m'}=\delta_{2p}^{(m-m'),0}\delta_{2n }^{(m+m'),0 }\eqno (4-5) $$
In this form we can better see another oversimplifying feature of $SU(2)$.
In fact Weyl group of $SU(2)$  is isomorphic to $Z_2$ and has only one
generator whose action on weights can be shoust by a minus sign.
Now from (4-5) it is apparent that:
$$P^{-m,-m'}=P^{m,m'}\eqno (4-6) $$
which is the same as property (3-17), holding for all modular
invariants of $SU(2)$.

Finally from (2-4) together with (4-6) we have:
$$M^{m,m'}=2(P^{m,m'}-P^{-m,m'})\eqno (4-7) $$
Dropping overall coefficient 2 and using (4-5) gives:
$$M^{m,m'}=\delta_{2p}^{(m-m'),0}\delta_{2n }^{(m+m'),0 }-
\delta_{2p}^{(m+m'),0}\delta_{2n }^{(m-m'),0 }\eqno (4-8) $$
This is the general form of all modular invariants of $SU(2)$.

 From (4-8)
we see that changing the role of $p$ and $n $ only changes the sign  the
modular invariant.
So the number of independent $M^{m,m'}$'s for a given $g$ is equal
to the number of the divisors of  $g$ divided by two  . This is half of
the number of $P^{m,m'}$'s Therefore without loss of generality we can always
assume  $p<n $.
Also we use $M_{g/p}$ to show the modular invariant specified by the
divisor $p$ of a defenite          height $g$.
 $p=1$ for all $g$'s gives diagonal partition functions
and $p=2$ for even $g'$s
(equivalent to the choice ${\bf b}=g\alpha^1$) leads
to $D$-partition functions of $SU(2)$. So, in order to find other  partition
functions
 we should consider modular invariants with $p>2$, in (4-8), and
combine them with each other (and  with $A-D$ series if necessary) to omit
negative terms completely.

We will show that such a process is doable only in a very few
cases. For this purpose we consider terms which are only present
in a given modular invariant, and not in any other modular invariant. We
call such terms as {\it {special terms}}.
As we will see, every modular invariant does have such special terms.

If a modular invariant has special terms with both
positive and negative signs, it is not possible to make a partition function
from it by means of combining it with other modular invariants.
The following theoshows that most of the modular invariants are of this form .

\vskip 0.5cm

{\bf  Theorem 4-1 }

{\it All modular invariants of $SU(2)$ with $p>2$ have both positive and
negative special terms except  five
cases}
$(g/p=12/3,18/3,24/4,30/3;30/5)$

Proof: Let us try to see if there are special terms in a given modular
invariant specified by  g and p as $M_{g/p}$.
The term $(m,m^\prime)= (n+p,n-p)$ for $p>2 $ belongs to $M_{g/p}$ with
positive sign and does not exist in any other modular invariant. To see this
suppose there is a $p^\prime$ with $ g=p^\prime n^\prime $such that
$M_{g/p^\prime}$ includes $(n+p,n-p)$ ... Using (4-8) we should either have
$$\left\{ \eqalign {2n&=2xn' \cr 2p&=2yp'\cr}\right. \eqno (4-9a) $$
or
$$\left\{ \eqalign {2n&=2xp' \cr 2p&=2yp'\cr}\right. \eqno (4-9b) $$
Both conditions imply $xy=1$ i.e the two modular invariants are the same .

The case for negative special term is more subtle . If a negative term
$(pt+n, pt-n)$ with integer t prime to n exists in $M_{g/p} $ it must be
special.
To see it suppose it also belongs to some $M_{g/p'}$ then (4-8) implies one of
the following conditions,
$$\left\{ \eqalign {n&=xn' \cr pt&=yp'\cr}\right. \eqno (4-10a), $$
$$\left\{ \eqalign {n&=xp' \cr pt&=yn'\cr}\right. \eqno (4-10b). $$
where x and y are positive integers . Both of the above relations imply $t=xy$
This means either $x=1 $ hence $p=p'$ and $n=n'$ or otherwise t and n have
common divisor x, in contradiction       to our assumption.

So we will investigate the conditions of existence of t such that $(pt+n,pt-n)$
is in $M_{g/p}$. From $0<m,m'<g$ we have $n<tp<n(p-1)$.

If $n=4k+r$ take $ t=2k+r'$ with $0<r,r'<4$. Then the condition becomes
$4k+r<(2k+r')p<(4k+1)(p-1)$. Since $p>2$ the l.h.s. is satisfied. The r.h.s is
however nontrivial and is
$$2k+r'<(2k+r-r')(p-1) \eqno (4-11)  $$
If this inequality holor some value of $p$ it will certainly holds for larger
values as well. So let us examine it for small values of $p$. The smallest $p$
is 3 which gives,
$$3r'<2(k+r) \eqno (4-12)$$
Let us choose $r'$ as following which ensures $[t,n]=1$,
$$\left\{ \eqalign { r&=0\ \ \ \  r'=1 \cr
                     r&=1\ \ \ \  r'=1 \cr
                     r&=2\ \ \ \  r'=3 \cr
                     r&=3\ \ \ \  r'=2 \cr } \right. \eqno (4-12) $$
Then the inequality implies respectively , $ k>3/2, k>1/2, k>5/2, k>0 $ .
Therefore only for the cases $ n=4 \ (r=0,r'=1),\  n=6 \ (r=2,r'=3) $ and $n=10
\ (r=2,r'=3) $ the inequality may fail. With the consideration that $p<n$ a
closer look shows that we are left only with the possibilities
$M_{12/3},M_{18/3},M_{24/4},M_{30/5},M_{30/3} $.
In all these cases the inequality is violated and the modular invariants do not
include negative special terms . This proves the theorem.

However in the case of $M_{24/4}$ the negative terms such as (2,14) are shared
by $M_{24/3}$ which itself includes special negative terms. Hence this
possibility is also ruled out and does not lead to a partition function. The
other four cases are given below.

$$M_{12/3}=\{ [\chi_1{\bar \chi }_7-
\chi_2{\bar \chi }_{10}+
\chi_5{\bar \chi }_{11}]+C.C.\}
 -\vert \chi_3\vert^2 +\vert \chi_4\vert^2 -
\vert \chi_6\vert^2 +\vert \chi_8\vert^2
-\vert \chi_9\vert^2 \eqno (4-13) $$

$$M_{18/3}=-\vert \chi_3-
\chi_9+\chi_{15}\vert^2\eqno (4-14) $$

$$\eqalign {M_{30/3}=\{ &[\chi_1{\bar \chi }_{19}
-\chi_2{\bar \chi }_{22}
+\chi_4{\bar \chi }_{16}
-\chi_5{\bar \chi }_{25}
+\chi_7{\bar \chi }_{13}
-\chi_8{\bar \chi }_{28}\cr
&+\chi_{11}{\bar \chi }_{29}
+\chi_{14}{\bar \chi }_{26}
+\chi_{17}{\bar \chi }_{23}]+C.C.\}
-\vert \chi_3\vert^2 -\vert \chi_6\vert^2
-\vert \chi_9\vert^2\cr &+\vert \chi_{10}\vert^2
-\vert \chi_{12}\vert^2 -\vert \chi_{18}\vert^2
+\ \chi_{20}\vert^2 -\vert \chi_{21}\vert^2
-\vert \chi_{24}\vert^2 -\vert \chi_{27}\vert^2\cr} \eqno (4-15) $$

$$\eqalign {M_{30/5}=\{ &[\chi_1{\bar \chi }_{11}
+\chi_2{\bar \chi }_{22}
-\chi_3{\bar \chi }_{27}
-\chi_4{\bar \chi }_{16}
+\chi_7{\bar \chi }_{17} +\chi_8{\bar \chi }_{28} \cr
&-\chi_9{\bar \chi }_{21}
+\chi_{13}{\bar \chi }_{23}
-\chi_{14}{\bar \chi }_{26}
+\chi_{19}{\bar \chi }_{29}!+C.C.\}-\vert \chi_5\vert^2 +\vert \chi_6\vert^2
\cr
&-\vert \chi_{10}\vert^2
+\vert \chi_{12}\vert^2 -\vert \chi_{15}\vert^2 +\vert \chi_{18}\vert^2
-\vert \chi_{20}\vert^2 +\vert \chi_{24}\vert^2
-\vert \chi_{25}\vert^2\cr} \eqno (4-16) $$

Miracolously what happens to $M_{24/4} $does not take place for these four
cases and the well known exceptional series of the $SU(2)_k$ is obtained,

$$E_6= M_{12/3}+D_{12}= \vert \chi_1+\chi_7\vert^2+
\vert \chi_4+\chi_8\vert^2+
\vert \chi_5+\chi_{11}\vert^2\eqno (4-17) $$

$$E_7=M_{18/3}+D_{18}=\vert \chi_1+\chi_{17}\vert^2
+\vert \chi_5+\chi_{13}\vert^2
+\vert \chi_7+\chi_{11}\vert^2 $$

$$+[(\chi_3+\chi_{15}{\bar \chi }_9
+C.C.]+\vert \chi_9\vert^2 \eqno (4-18) $$

$$E_8= M_{30/3}+M_{30/5}+D_{30}=\vert \chi_1+\chi_{11}
+\chi_{19}+\chi_{29}\vert^2
+\vert \chi_7+\chi_{13}
+\chi_{17}+\chi_{23}\vert^2 \eqno (4-19) $$

Now we have completed the derivation of the complete $SU(2)$ series using only
the mono-shift method. With the use of the point that bosonic modular invariant
partition functions are obtained by mono-shifts this constitutes another proof
of the A-D-E classification.

\vskip 10pt

{\bf 5-Summary and Disscussion}
\vskip 5pt

What we presented in this paper was a natural exploration
 of the relation  between modular invariants of
$\hat G $ and  $\hat H$-theories. There are two basic quantities in this
relation, namely: supersymmetry index $I_{\lambda }^{\Lambda }$ and (its
subalgebra) $\hat H$-moduinvariant  $P^{\lambda ,\lambda '} $ ,
 we have given explicit formulas for each of them. For
$I_{\lambda }^{\Lambda }$, this is simply done  (2-3).
But for $P^{\lambda ,\lambda '}$ it takes a lengthy calculation
which leads to (2-20) or (2-25).

We found that the only relevant part of shift vector is the difference between
left and right parts of it, which can be specified by a weight ${\bf b}\in
{\Gamma ^\ast (G)\over {g\Gamma (G)}}$
(see 2-11)
hence the number of resulted modular invariants it much less than what
seems in the beginning.
All characteristic quantities of each modular invariant
can be extracted from the shift vector $\bf b$.

We introduced the concept of  associated weights,  which,  is characterized by
a
simple inhomogeneous linear relation between weights specifying left and
right representations in different terms of $\hat H$-modular invariants (see
2-34). Moreover, we observed that the structure of $\hat H $ modular invariants
is such that definite $\lambda $-member families of characters in the left and
right
sectors couple to each other (see 2-38).

 The main  technical difficulty in our approach is
 the double summation over Weyl group of G in (2-4), which prevents
us to
get a simple relation between left and right representations taking part in a
modular invariant.
As mentioned in the text, in fact only one of
the summations is the real one, and in the other summation only one unique
element of $ W(G)$ takes part.

There can also be found special circumstances which we can remove this
difficulty. That is e.g.  when $P^{\lambda ,\lambda '}$ is invariant under the
action
of weyl
group of $G$ (see 3-20), all terms in the summation over $ W(G)$ give the same
result,
leading to an overall $\vert W (G) \vert $ factor in the final expression
(which
can be dropped).
In this work we introduced two such special situations. The f one occurs
when we choose the shift vector $\bf b$ in the lattice $g\Gamma^\ast (G)$ . As
shown in the
text, independent results of such choices lead to the partition
functions of the $D$-series. We followed the procedure for affine $SU(N) $ at
arbitrary level, but similar treatment can be done for all other simple Lie
groups.

The second situation  where we encounter the invariance of $ P^{ \lambda,
\lambda' }$
under $W(G)$, is affine $SU(2)$ at arbitrary level. In this case, the Weyl
group is  isomorphic to $Z_2$ and the invariance holds naturally.
Another special feature of $SU(2)$ is that it is a rank-1 group and
mono-shift method gives all of its modular invariants. So we are able to go
through the
classification of its modular invariants and partition functions, on the basis
of
our method. In this way we find a new proof of the A-d-E classification.

We conjecture that  our method  can be generalized to higher rank affine
algebras is like $SU(3)_k$, etc. where we need shifts with more steps whose
number increases with the rank of the algebra to ensure covering of space of
modular invariants . We only have to detect simultaneous presence of positive
and negative terms to rule out a modular invariant.

Our experience with the $SU(2)$ case shows that positivity is a very strong
restriction and one expects that few exceptional cases may exist for higher
rank
algebras where the inequality $0<m<g$ is replaced by $ 0< \sum m_i<g$. Work
along this direction is in progress .

After this work was written we received the reference 14 where the complete
classification of the modular invariant partition functions of $SU(3)$ is
given. Also a new proof of the A-D-E classification is worked out . They
emphasis    on the positivity from the beginning . Their
method is different from ours. They show that for the $SU(3)$ case all the
possible partitionctions are the ones that are already known.
\vskip 10pt

{\bf Acknowledgements}:

We are grateful to F.Ardalan, W.Nahm , S.Rouhani, C.Vafa and M.Abolhassani for
useful discuss        ions. We also thank F. Ardalan for his encouragements and
participation in the early stages of the work. H.A wishes to thank Max-Planck
Institute for Mathematics for hospitality and financial support.
\vfill \break

 \centerline {\bf {Appendix A: Proof of Theorem (2-2)} }
\vskip 0.5cm

Suppose $\bar \lambda $ and ${\bar \lambda }'$ are associated weights to
$\lambda $ due to
shift vectors $\bf b$ and $r{\bf b}$ respectively. From (2-24) we have
$$\left\{ \eqalign {&{\bar \lambda }-\lambda =j{\bf b}\cr
&({\bar \lambda }+\lambda ).\beta =2kn \cr}\right.;
\qquad \left\{ \eqalign {&{\bar \lambda }'-\lambda =j'r{\bf b}\cr
&({\bar \lambda }'+\lambda ).r\beta =2k'n }\right. \eqno (A-1)$$
for some integers $j,k,j'$ and $k'$. Subtracting similar equations in (A-1)
from
each other gives:
$$\left\{ \eqalign {&{\bar \lambda }'-{\bar \lambda }=(j'r-j){\bf b}\cr
(&{\bar \lambda }'-{\bar \lambda }).r\beta =2(k'-rk)n }\right. \eqno (A-2)$$
which yields the following equation between integers involved:
$$(k'-rk)n '=(j'r-j)rt'\eqno (A-3) $$
In this integer equation, l.h.s. is a multiple of $n '$, but in r.h.s. $r$ and
$t'$
are both prime relative to $n '$ ($[r,n]=1$ because $[r,qg]=1$). So $(j'r-j)$
should
be a multiple of $n '$, which gives:
$${\bar \lambda }'-{\bar \lambda }=0\pmod {n 'b}\eqno (A-4) $$

This shows that ${\bar \lambda }'$ belongs to the set of ${\bar \lambda
}_{\imath }$'s resulted
from shift vector $\bf b$. It should also be added that $\mu =t,n !$ is the
same for
both modular invariants, and since $r,n !=1$, cyclic group
of powers of operator $\xi $ is also the same. So belonging
${\bar \lambda }'$ to the set of ${\bar \lambda }_{\imath }$'s requires
identifica of associated
families to $\lambda $ in both modular invariants.

\vskip 1cm

\centerline {\bf {Appendix B: Completeness  of Monoshift Method  for $SU(2)$}}

\vskip 0.5cm

An  even self-dual lorentzian lattice ${\cal L}'$ of dimension (1,1)
which contains the sublattice $\bigl(\ \sqrt g\Gamma (SU(2)),\sqrt g\Gamma
(SU(2))\ \bigr)$  consists of elements like {\bf v} given below  in the basis
$\bf \alpha^1/\sqrt g$ for left and right parts of each vector, an arbitrary
element of
$${\bf v}=(m+2gi,m'+2gi')
\eqno (B-1) $$
where $i$ and $i'$ are arbitrary integers and $m$ and $m'$ are integers in
the interval $[0,2g-1]$ and satisfying the condition:
$$m^2-{m'}^2=0\pmod {4g}\eqno (B-2a)$$

We shall show that elements of such a lattice has the form 4-5 and hence prove
that all modular invariants of $SU(2)$ is obtained by one shift of the diagonal
lattice.

 If $\bf v_1$ and
$\bf v_2$ are two arbitrary vectors in ${\cal L}'$ specified by pairs $(m_1,
m'_1)$ and
$(m_2, m'_2)$ respectively, they should also satisfy the constraint:
$$m_1m_2-{m'}_1{m'}_2=0\pmod {2g} \eqno (B-2b) $$
 From (B-2a) we observe that $m$ and $m'$ should be both even or both odd,
and $g$ should be decomposed into two factors $p$ and $n $ in such a way that
$$\left\{ \eqalign {m-m'&=0\pmod {2p}\cr m+m'&=0\pmod {2n }\cr}\right. \eqno
(B-3)$$
Suppose we have decompositions $g=p_1n_1 $ for $(m_1,{m'}_1)$
and $g=p_2{n }_2$ for $(m_2,{m'}_2)$, such that:
$$\left\{ \eqalign {m_1-{m'}_1&=2x_1p_1\cr m_1+{m'}_1&=2y_1{n }_1\cr}\right.;
\qquad \left\{ \eqalign {m_2-{m'}_2&=2x_2p_2\cr m_2+{m'}_2&=2y_2{n
}_2\cr}\right. \eqno (B-4) $$

Assume $q $ is the g.c.d of $p_1 $ and $p_2$such that $p_1=qr_1$,$p_2=qr_2$.
Then clearly
$g=qur_1r_2,\  \ , n_1=ur_2 $ and $n_2=ur_1$ and we have

$$\left\{ \eqalign {m_1-{m'}_1&=2x_1qr_1\cr m_1+{m'}_1&=2y_1{ur_2 }\cr}\right.;
\qquad \left\{ \eqalign {m_2-{m'}_2_2qr_2\cr m_2+{m'}_2&=2y_2{ur_1
}\cr}\right. \eqno (B-5) $$

For these relations to be nontrivial we must have
$[x_1,r_2]=[x_2,r_1]=[y_1,r_1]=[y_2,r_2]=1$.
Condition (B-2b)implies that
$$x_1y_2r_1+x_2y_1r_2=0 \ mod(r_1r_2)  \eqno (B-6)$$
Hence using $[r_1,r_2]=1$ we deduce that $y_1=s_1r_1 ,\  y_2=s_2r_2 $ with
$s_1$
and $s_2$  integers. Then we have

$$\left\{ \eqalign {m_1-{m'}_1&=x_1qr_1\cr m_1+{m'}_1&=2s_1r_1ur_2 \cr}\right.;
\qquad \left\{ \eqalign {m_2-{m'}_2&=2x_2qr_2\cr m_2+{m'}_2&=2s_2r_1ur_2
\cr}\right. \eqno (B-7) $$
It is clear now that both elements are in the space obtained by taking p to be
equal to q the common divisor of $p_1$ and $p_2$ and n equal to $r_1ur_2$ in
(4-5).

\vfill \break

\vskip 0.5cm

{\bf {References}}

\vskip 0.5cm

\item {1.} J. L. Cardy, Nucl. Phys. B270 {FS16! (1986) 186.
\item {2.} D. Gepner and E. Witten, Nucl. Phys. B278 (1986) 493.
\item {3.} P. Bouwgknet and W. Nahm, Phys. Lett. B184 (1987) 359.
\item {4.} C. Ahn and M. A. Walton, Phys. Lett. B223 (1989) 343;
F.Ardalan, H. Arfaei and S. Rouhani, Int. Jour. Mod. Phys. A26 (1991) 4763.
\item {5.} A. Schellekens and S.Yankilowicz,Nucl. Phys. B327 (1989) 673;
Nucl. Phys. B334 (1990) 67.
\item {6.} M. Bauer and C. Itzykson, Commun. Math. Phys. 127 (1990) 617.
\item {7.} A. Capelli, C. Itzykson, and J. B. Zuber, Nucl. Phys. B280
(1987) 445; Commun. Math. Phys. 113 (1987).
\item {8.} A. Kato, Mod. Phys. Lett. A8 (1989) 585.
\item {9.} P.Degiovanni, Commun. Math. Phys. 127 (1990) 71
\item {10.} Ph. Ruelle,E.Thiran and J.Weyers, Commun. Math. Phys. 133 (1990)
305
\item {11.} N. P. Warner, Commun. Math. Phys. 130 (1990) 205.
\item {12.} W. Lerche, C. Vafa and N. P. Warner, Nucl. Phys. B324 (1987) 427
\item {13.} K. S. Narain, M. H. Sarmadi, and E. Witten, Nucl. Phys. B279
(1986) 369; K. S. Narain, Phys. Lett. B169 (1986) 41.
\item {14.} T. Gannon, The clfication of affine $SU(3)$ Modular Invariant
Partition Functions ,preprint , Mathematics Department, Carlton University,
Ottawa, Ontario,Canada.
\end